# Mechanism of anatase-to-columbite $TiO_2$ phase transformation via sheared phases: first-principles calculations and high-pressure torsion experiments

Jacqueline Hidalgo-Jiménez[1,2], Taner Akbay[3], Yuji Ikeda[4], Tatsumi Ishihara[1,5,6] and Kaveh Edalati[1,2,6,*]

[1] WPI, International Institute for Carbon-Neutral Energy Research (WPI-I2CNER), Kyushu University, Fukuoka, Japan
[2] Graduate School of Integrated Frontier Sciences, Department of Automotive Science, Kyushu University, Fukuoka, Japan
[3] Department of Materials Science and Nanotechnology Engineering, Yeditepe University, Istanbul, Turkey
[4] Department of Materials Design, Institute for Materials Science, University of Stuttgart, Stuttgart, Germany.
[5] Department of Applied Chemistry, Faculty of Engineering, Kyushu University, Fukuoka, Japan
[6] Mitsui Chemicals, Inc. - Carbon Neutral Research Center (MCI-CNRC), Kyushu University, Fukuoka, Japan

**Abstract**

High-pressure torsion (HPT) can facilitate phase transformations in titanium dioxide ($TiO_2$) and stabilize its high-pressure columbite phase, as an active photocatalyst, by shear straining under high pressure. This study aims to understand the mechanism underlying the acceleration of the anatase-to-columbite phase transformation by shear strain. A mechanism by considering sheared crystal structures as intermediate phases was proposed and examined using quantum mechanics in the framework of density functional theory (DFT) and HPT experiments. DFT energy and phonon calculations demonstrated the viability of the sheared structures as intermediate phases. Furthermore, the sheared structures were observed experimentally as new metastable phases using high-resolution transmission electron microscopy. These findings can explain the significant effect of shear strain on pressure-induced phase transitions, reported during severe plastic deformation of various metals and ceramics.

*Corresponding author (E-mail: kaveh.edalati@kyudai.jp; Tel: +81-92-802-6744)

## 1. Introduction

Severe plastic deformation methods demonstrate outstanding capability to modify the crystal structure and microstructure of different types of materials to enhance mechanical and functional properties [1,2]. Particularly, high-pressure torsion (HPT) is a unique technique applicable to almost any kind of materials, in which high pressure and torsional shear strain are simultaneously applied to a small workpiece [3,4]. In the HPT method, as shown in Fig. 1(a), the sample is placed between two anvils, compressed under high pressure, and subsequently exposed to extremely large shear strain by rotation [3,4]. A relationship to calculate the shear strain ($\gamma$) during HPT is given in equation 1, where $r$ is the radial distance from the rotation center, $N$ is the number of rotations and $h$ is the sample thickness [1,4].

$$\gamma = \frac{2\pi r N}{h} \qquad (1)$$

The severe conditions in HPT force the materials, even hard and brittle ones such as oxides and ceramics [5,6], to microstructural changes and grain refinement as well as to structural changes and phase transformations [1-4]. The microstructural evolution involves the presence of defects like vacancies and dislocations and ultrafine/nano-grain formation [7-9]. However, after a certain number of rotations, the induced strain does not generate more changes due to reaching a dynamic equilibrium [10,11]. The dynamic equilibrium implies that not only does the rate of defect production and annihilation reach a steady state but also phase transformations stop [10,11]. Shear strain during HPT reduces the apparent pressure required for phase transformations and leads to the detection of hidden intermediate phases in some metals and ceramics [11-13].

The HPT method has been applied to a large variety of ceramics to study changes in phase transformations. $Al_2O_3$ [6,14], $ZrO_2$ [15], ZnO [16], $Y_2O_3$ [17], $BaTiO_3$ [6] and $TiO_2$ [18] are some examples of ceramics that show phase transformation after the application of HPT. For $TiO_2$, HPT was able to stabilize the high-pressure columbite phase which shows good photocatalytic properties [19-22]. Fig. 1 (b) shows the phase diagram for $TiO_2$ which includes anatase (space group: $I4_1/amd$), rutile (space group: $P4_2/mnm$) and columbite (space group: $Pbcn$) [23]. As shown in the phase diagram, columbite can be formed at around 2 GPa at room temperature [18,23], but the stabilization of this phase under ambient pressure is challenging. Previous studies demonstrated that shear strain induced by HPT positively influences the formation and stability of the columbite phase [19-22]. However, the reason for the impact of shear strain on the formation and stability of the columbite phase has not been well understood yet.

This study thus aims to explain the effect of shear strain on the anatase-to-columbite transformation mechanism by the formation of intermediate sheared phases during the HPT process. For this reason, quantum mechanics calculations and experiments using the HPT method are employed.

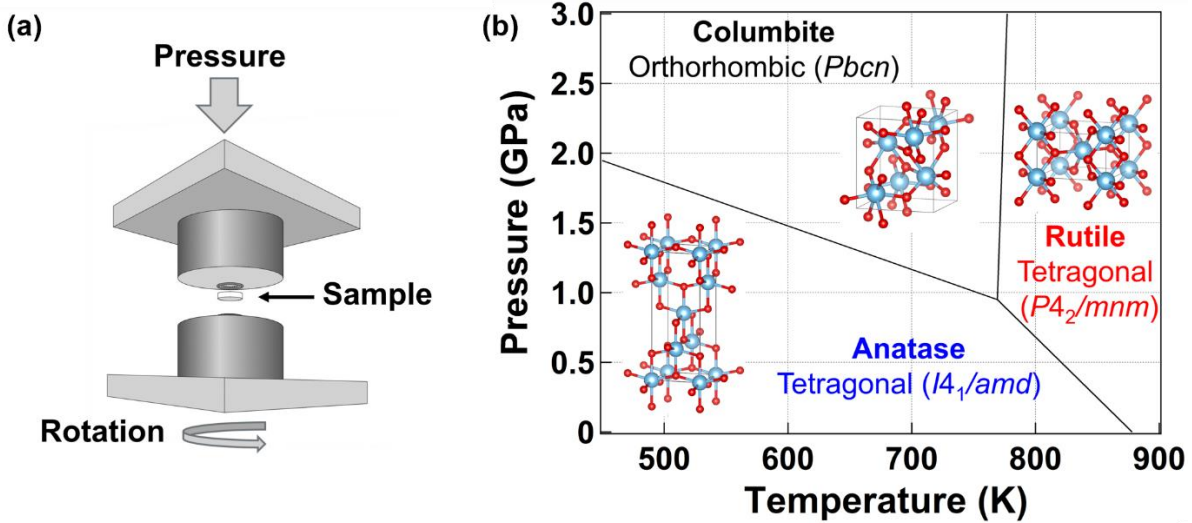

Fig. 1. (a) Schematic representation of HPT process and (b) $TiO_2$ pressure-temperature phase diagram.

## 2. Materials and Methods
### 2.1. Computational Methods

Several models were developed to mimic the HPT conditions to understand the effect of shear strain on phase transformations. The models were developed by shifting the (001) plane along the *b* axis so that the shear strain on the structures would be $\gamma = 0.1$, 0.2, and 0.4. Fig. 2 shows the two-dimensional representation of the shear strain of 0.1 over the unit cells of anatase, rutile and columbite. The unsheared (indicated as pristine, hereafter) and sheared models were studied using first-principles calculations by employing density functional theory (DFT) in the Vienna *Ab-initio* Simulation Package (VASP) [24,25]. Generalized gradient approximation (GGA) and Perdew-Burke-Ernzerhof (PBE) functional [26] were utilized to describe the exchange correlation. Projected-Augmented Wave (PAW) [27] of Ti_sv and O provided by VASP were used for titanium and oxygen atoms, respectively. The cutoff for the plane-wave functions was set at 520 eV. The Brillouin zone was sampled with a k-points grid of 8×8×6 for anatase and 8×8×8 for rutile and columbite considering the geometry of the structures. The tetrahedron method was implemented for Brillouin zone integrations [28]. The crystallographic information of pristine anatase, rutile and columbite was taken from Materials Project, and optimized under ambient pressure and 6 GPa (PSTRESS = 60) allowing the mobility of positions, cell shapes, and cell volumes (ISIF = 3). Subsequently, the shear was applied, and all ion positions were optimized again by constraining the cell shapes and volumes (ISIF = 2). The crystallographic information for pristine anatase, rutile, and columbite optimized under ambient pressure are compared with the experimental information in Table 1, indicating good agreement between the calculated and experimental values. The characteristics of all the sheared crystal structures after optimization are also given in Table 2. The electronic and ionic constraints of $1.0\times10^{-6}$ and $1.0\times10^{-5}$ were applied, respectively. A Hubbard $U$ parameter [29] equal to 8 eV was set for titanium atoms in the calculations after several tests which consistently agreed with the literature [30]. Furthermore, density functional perturbation theory (DFPT) was utilized for phonon calculations and analyzing the dynamic stability of the structures using the code Phonopy [31,32]. The supercell size employed was 2×2×2 of the initial unit cells.

Table 1. Comparison of crystallographic information of pristine anatase, columbite and rutile with corresponding experimental values under ambient pressure.

| Structure | Space Group | Method | Angles (º) $\alpha = \beta = \gamma$ | Lattice Parameters (Ambient Pressure) $a$ | $b$ | $c$ | JCPDS Number |
|---|---|---|---|---|---|---|---|
| Anatase | $I4_1/amd$ | Calculation | 90 | 3.90 | 3.90 | 9.82 | |
| | | Experiment | 90 | 3.78 | 3.78 | 9.5 | 01-084-1285 |
| Rutile | $P4_2/mnm$ | Calculation | 90 | 4.65 | 4.65 | 2.96 | |
| | | Experiment | 90 | 4.60 | 4.60 | 2.97 | 01-076-1939 |
| Columbite | $Pbcn$ | Calculation | 90 | 4.58 | 4.94 | 5.58 | |
| | | Experiment | 90 | 4.53 | 4.90 | 5.50 | 01-084-1750 |

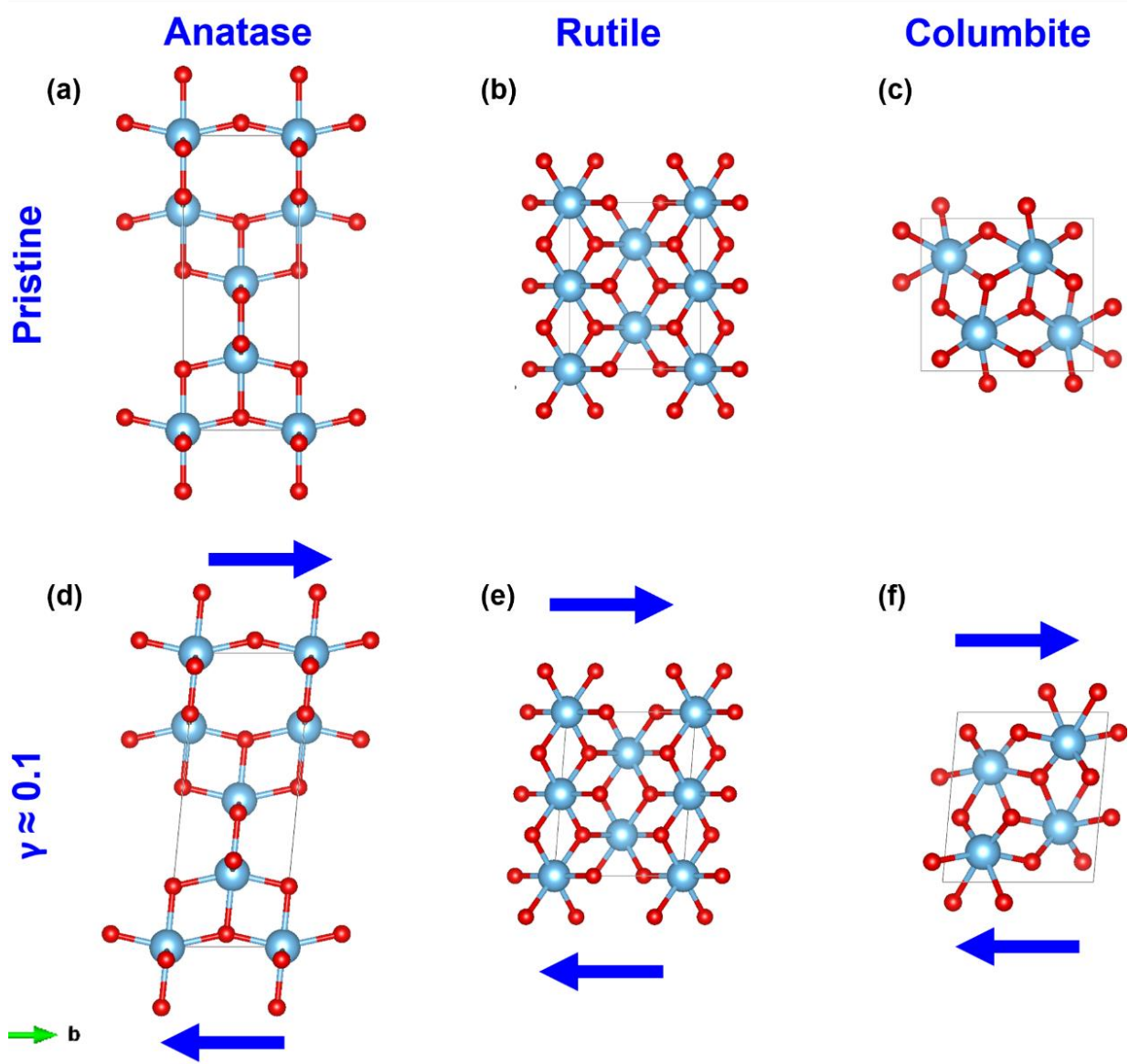

Fig. 2. Unit cell models of (a, d) anatase, (b, e) rutile and (c, f) columbite for (a-c) pristine structures and (d-f) structures sheared with $\gamma = 0.1$ along $b$ axis.

Table 2. Crystallographic information of pristine and sheared structures of anatase, columbite and rutile. Sheared structures were achieved by uniaxial shearing of pristine structures along *b* axis while keeping volume constant.

| Structure | Space Group | Angles (°) | | Lattice Parameters (Å) at Ambient Pressure | | | Lattice Parameters (Å) at 6 GPa | | |
|---|---|---|---|---|---|---|---|---|---|
| | | $\alpha$ | $\beta = \gamma$ | a | b | c | a | b | c |
| **Anatase** | $I4_1/amd$ | 90.00 | 90.00 | 3.90 | 3.90 | 9.82 | 3.87 | 3.87 | 9.62 |
| **Anatase, $\gamma = 0.1$** | $C2/m$ | 84.41 | 90.00 | 3.90 | 3.90 | 9.86 | 3.87 | 3.87 | 9.67 |
| **Anatase, $\gamma = 0.2$** | $C2/m$ | 78.93 | 90.00 | 3.90 | 3.90 | 10.00 | 3.87 | 3.87 | 9.80 |
| **Anatase, $\gamma = 0.4$** | $C2/m$ | 68.69 | 90.00 | 3.90 | 3.90 | 10.53 | 3.87 | 3.87 | 10.32 |
| **Rutile** | $P4_2/mnm$ | 90.00 | 90.00 | 4.65 | 4.65 | 2.96 | 4.64 | 4.64 | 3.05 |
| **Rutile, $\gamma = 0.1$** | $P2_1/c$ | 84.20 | 90.00 | 4.65 | 4.65 | 2.98 | 4.64 | 4.64 | 3.07 |
| **Rutile, $\gamma = 0.2$** | $P2_1/c$ | 78.92 | 90.00 | 4.65 | 4.65 | 3.03 | 4.64 | 4.64 | 3.11 |
| **Rutile, $\gamma = 0.4$** | $P2_1/c$ | 68.61 | 90.00 | 4.65 | 4.65 | 3.19 | 4.64 | 4.64 | 3.28 |
| **Columbite** | $Pbcn$ | 90.00 | 90.00 | 4.58 | 4.94 | 5.58 | 4.57 | 5.06 | 5.57 |
| **Columbite, $\gamma = 0.1$** | $P2_1/c$ | 84.95 | 90.00 | 4.58 | 4.94 | 5.60 | 4.57 | 5.06 | 5.60 |
| **Columbite, $\gamma = 0.2$** | $P2_1/c$ | 78.80 | 90.00 | 4.58 | 4.94 | 5.71 | 4.57 | 5.06 | 5.68 |
| **Columbite, $\gamma = 0.4$** | $C2/c$ | 68.20 | 90.00 | 4.58 | 4.94 | 6.10 | 4.57 | 5.06 | 6.00 |

**2.2. Experimental Methods**

To examine the formation of columbite, HPT in quasi-constrained conditions was used to process the sample. The HPT anvils were made of a composite of tungsten carbide with 11 wt% cobalt and had a cylindrical shape (50 mm diameter and 50 mm height) with a flat-bottomed hole on the surface (10 mm diameter and 0.25 mm depth). Five disc-shaped pellets were generated by pressing ~260 mg of anatase powder (99.8%) under 30 kN. Subsequently, each pellet was subjected to HPT under different conditions: compression under 6 GPa at (1) 523 K, (2) 573 K, and (3) 623 K, and HPT processing under 6 GPa at room temperature for (4) 3 turns and (5) 15 turns. The resulting materials had the shape of a disc with 10 mm diameter and ~0.8 mm thickness. The outer parts of discs ($r > 2$ mm) were crushed with a mortar and pestle and examined by X-ray diffraction (XRD) using Cu Kα radiation and transmission electron microscopy (TEM) with an acceleration voltage of 200 keV.

## 3. Results

### 3.1. Computational Results

To understand the anatase-to-columbite transformation pathway, the suggested sheared structures were examined by DFT calculations. Self-consistent field calculations (SCF) were performed on pristine and sheared models to understand their thermodynamic stability. The enthalpies (referenced to that of anatase at ambient pressure without shearing) are plotted in Fig. 3. At ambient pressure (Fig. 3a), the enthalpies of all the phases increase after applying the shear strain, and thus the structures become thermodynamically less stable compared to pristine structures, as expected. Moreover, the enthalpy differences between pristine columbite, anatase and rutile are quite small, suggesting a possible reason for the reported metastability of columbite at ambient conditions [19-22]. In other words, the driving force for the transformation of columbite to rutile (the most stable phase of $TiO_2$ [33]) is in the range of 0.01 eV, suggesting that this phase transition cannot happen without the introduction of thermal energy similar to anatase-to-rutile phase transformation. Since the HPT process was carried out under large pressures such as 6 GPa, the calculations were also performed under a pressure of 6 GPa, as shown in Fig. 3b. The most important point that can be observed in Fig. 3b is that the stability of sheard phases compared to anatase changes by increasing pressure to 6 GPa. While all sheared phases have higher energy than anatase under ambient pressure, the phases sheared with $\gamma = 0.1$ have similar or smaller energy levels than anatase under 6 GPa. These energy differences indicate that the transformation of anatase to phases sheared with $\gamma = 0.1$ is energetically favorable under HPT processing conditions (i.e. under 6 GPa with simultaneous shear strain application). Two points can be derived from Fig. 3b: (i) it is more likely to observe rutile and columbite under 6 GPa, regardless of shearing, and (ii) a possible mechanism can be suggested for the transformation of anatase to columbite and/or rutile through the formation of intermediate sheared phases. Such possible intermediate phases can be of interest not only from the phase transformation point of view but also for controlling the properties of materials [34,35].

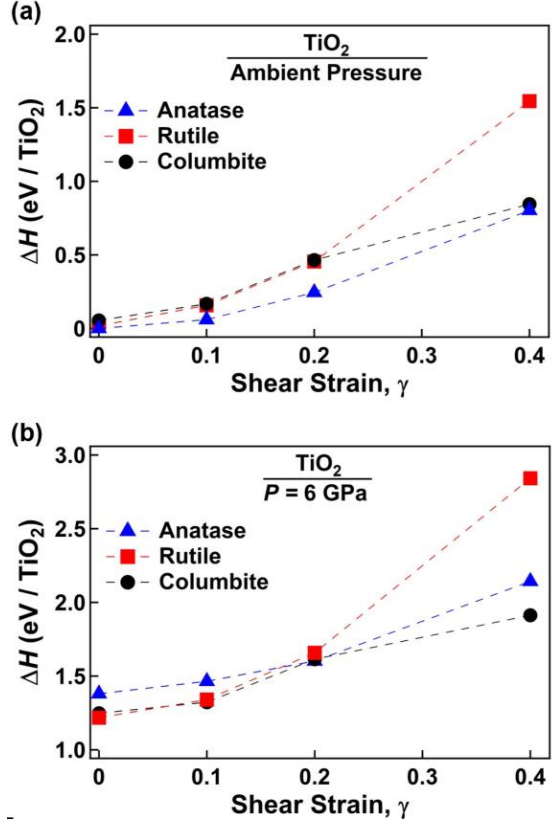

Fig. 3. Variation of energy differences (ΔH) obtained from self-consistent field calculations for pristine and sheared polymorphs under (a) ambient pressure and (b) $P = 6$ GPa, using pristine anatase in ambient pressure as the ground state.

Considering the proposed mechanism, the dynamic stability of anatase, rutile and columbite without and with a shear strain of $\gamma = 0.1$ was studied. Fig. 4 shows the phonon calculations for the pristine and sheared structures. For the pristine structures (Fig. 3a, 3c and 3e), none of them show any imaginary modes which suggests that the three polymorphs are dynamically stable under ambient pressure. On the other hand, when analyzing the sheared structures, only anatase and columbite sheared structures (Fig. 3b, 3f) are dynamically stable and do not show imaginary modes [31]. The sheared rutile (Fig. 3e) shows imaginary modes that can be interpreted as its instability. Since the anatase and columbite sheared phases are dynamically stable, the mechanism of phase transformation from anatase to columbite under pressure and shear conditions might involve the presence of distorted phases going from pristine anatase → sheared anatase → sheard columbite → pristine columbite. Although such a mechanism has not been observed for TiO$_2$ so far, some theoretical and experimental studies already mentioned the possibility of the formation of hidden phases under high pressures and concurrent stain for some other materials [12,13].

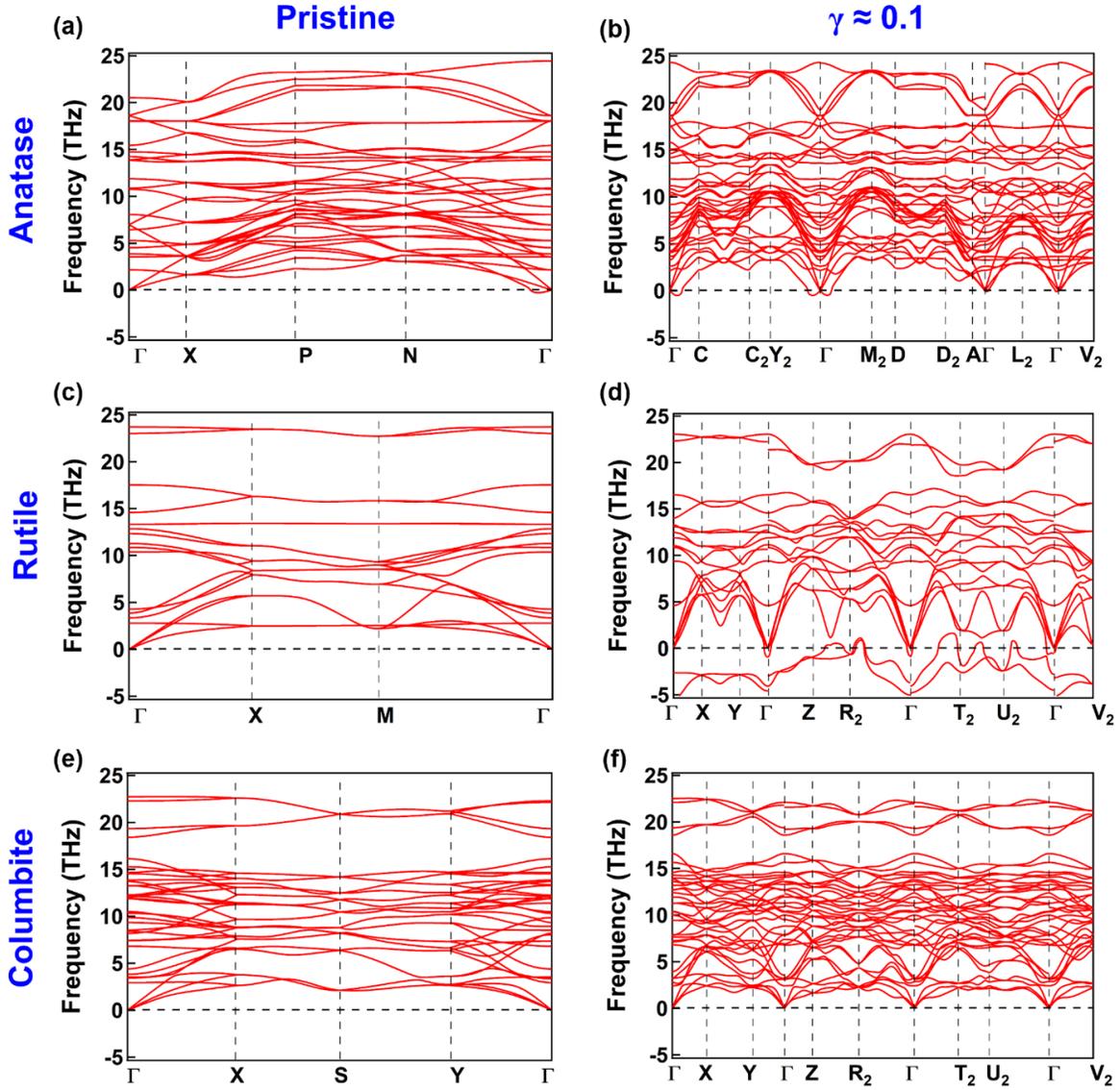

Fig. 4. Phonon spectra of (a, b) anatase, (c, d) rutile and (e, f) columbite for (a, c, e) pristine and (b, d, f) sheared structures with $\gamma = 0.1$, calculated by density functional perturbation theory. Imaginary modes are shown in negative values.

## 3.2. Experimental Results

As mentioned above, HPT can make multiple changes in the microstructure of materials and phase transformation is one of these changes [1-4]. Fig. 5 shows the XRD patterns for anatase processed by (a) pure compression and (b) HPT. A comparison of these profiles indicates several important points. First, the samples processed under 6 GPa by compression and HPT processing show the appearance of new peaks for the high-pressure columbite phase. Second, in samples processed by pure compression, the intensity of peaks for the columbite phase slightly increases with increasing the compression temperature. It should be noted that no columbite peak could be detected by compression at room temperature. Third, the fraction of the columbite phase significantly increases with the application of shear strain through the HPT process, and the maximum columbite fraction (~70 wt%) is achieved for the sample processed by HPT for 15 turns. It is then concluded that shear strain is the most significant parameter in facilitating the anatase-to-columbite phase transformation [18-22]. Fourth, a peak broadening occurs after HPT processing (Fig. 5b), which is an indication of crystallite size reduction and linear defect formation [7,36].

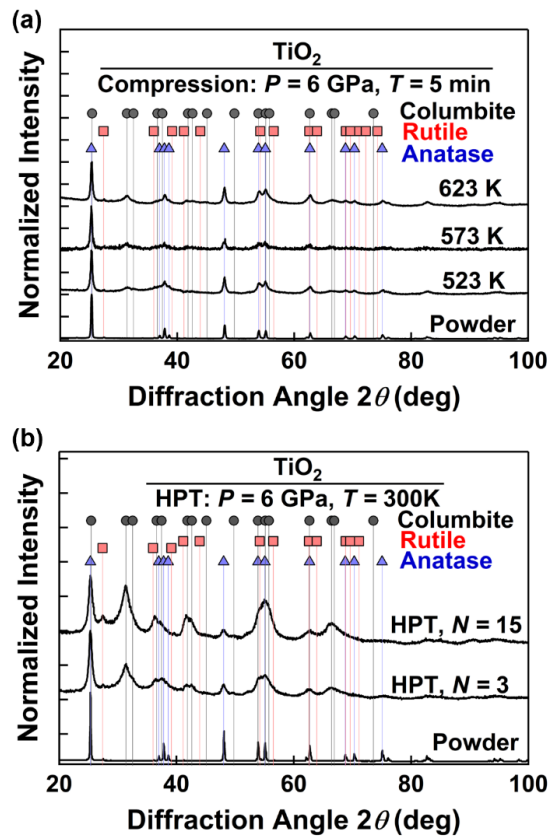

Fig. 5. XRD patterns for anatase powders and for samples processed under 6 GPa by (a) compression at temperatures of 523, 573 and 623 K and (b) HPT at room temperature for 3 and 15 turns.

A magnified view of the XRD pattern of the HPT-processed sample after 15 turns is shown in Fig. 6a. In addition to the formation of the columbite phase, peaks for the rutile phase are visible in Fig. 6a. Here, three features can be mentioned. First, for the anatase standard XRD profile, the peak located at 25.3° should have an intensity 4 times higher than the one located at 48.0°. However, the intensity of this peak in Fig. 6a is 7 times larger than the one at 48.0°. This indicates that the peak at 25.3° should correspond not only to anatase but also to other phases. Second, for the rutile standard peak profile, the peak located at 36.0° should have an intensity 2 times higher than the peak located at 27.4°. However, the intensity ratio for these two peaks is the inverse of the standard peak ratio. This indicates that the peak at 36.0° corresponds not only to rutile but also to other phases. Third, there is a small shoulder at ~52° which can be explained neither by anatase, rutile nor columbite. One explanation for these three features can be the presence of intermediate sheared phases. Fig. 6b-d shows the calculated XRD patterns for the sheared phases, illustrated in Fig. 2. These calculated XRD patterns suggest that the three features observed in the HPT-processed sample can be due to the presence of sheared phases, as compared using arrows in Fig. 6. Here, it should be noted that because of the broadness of XRD profile and the presence of various phases, it was not possible to conduct a reliable Rietveld refinement and quantify the fraction of sheared phases [37]. However, from peak intensities, it is expected that the fraction of these sheared phases should be quite low. Moreover, the sheared rutile phase is not expected to exist because it is dynamically unstable, as discussed using phonon calculations in Fig. 4 [31].

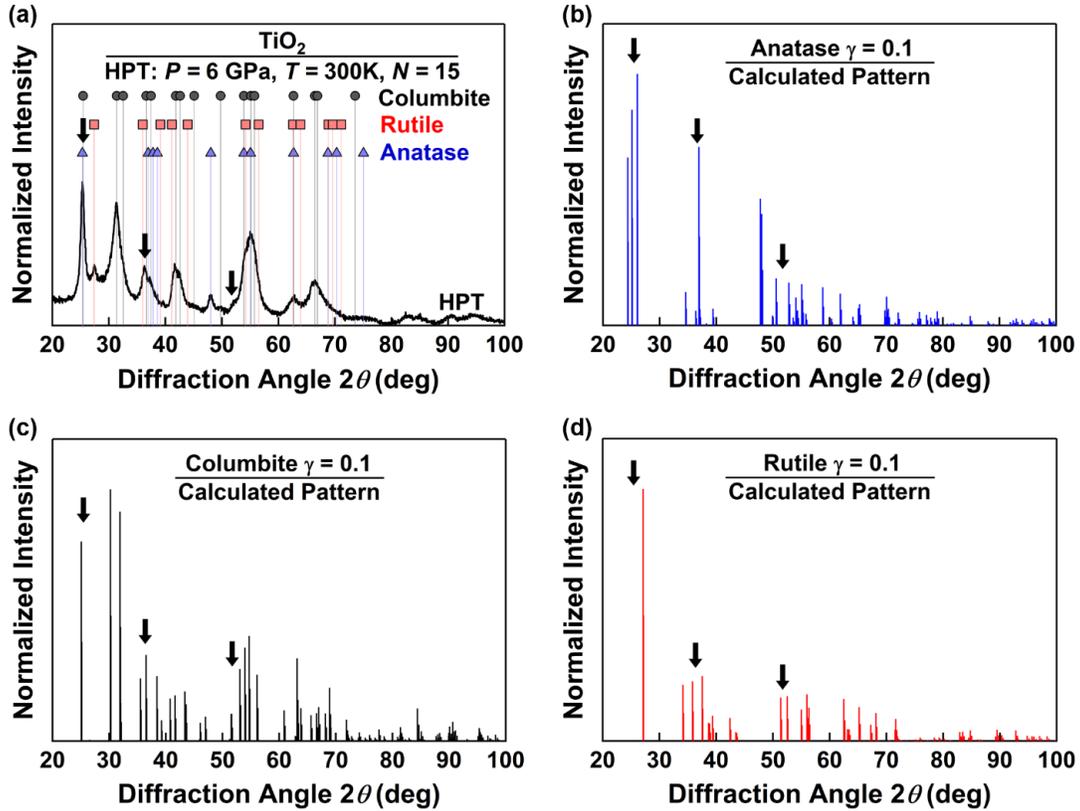

Fig. 6. (a) XRD patterns of sample processed by HPT under 6 GPa at room temperature for 15 turns in comparison with XRD profiles of sheared phases of (b) anatase, (c) columbite and (d) rutile with $\gamma = 0.1$.

In an attempt to find another piece of evidence for the presence of sheared phases, over 100 lattice images were examined by high-resolution TEM and fast Fourier transform (FFT) diffractograms for the sample processed by HPT for 15 turns. Fig. 7 shows four representative high-resolution images. Detailed analysis of these images confirms the presence of anatase, rutile and columbite phases in good agreement with the XRD profile of Fig. 6a. However, some lattice images were detected that could not be identified as anatase, rutile or columbite. A comparison between the FFT diffractograms of these lattice images with the calculated diffractograms for sheared phases confirms that they correspond to the sheared anatase phase with $\gamma = 0.1$, as shown for one crystal in Fig. 7a and another one in Fig. 7c. Among over 100 examined lattice images, no evidence for the presence of sheared columbite and rutile phases could be found. The absence of sheared rutile phase may be due to its dynamical instability. However, the absence of the sheared columbite phase indicates that either it is not an intermediate phase or its fraction is low. Since

~100 lattice images were analyzed in this study, the absence of one lattice image of the sheared columbite phase statistically explains that the fraction of this phase (if any) should be less than 1/100 or 1%. It should be noted that the shear anatase phase has not been reported so far in the literature, perhaps because of its metastability [23].

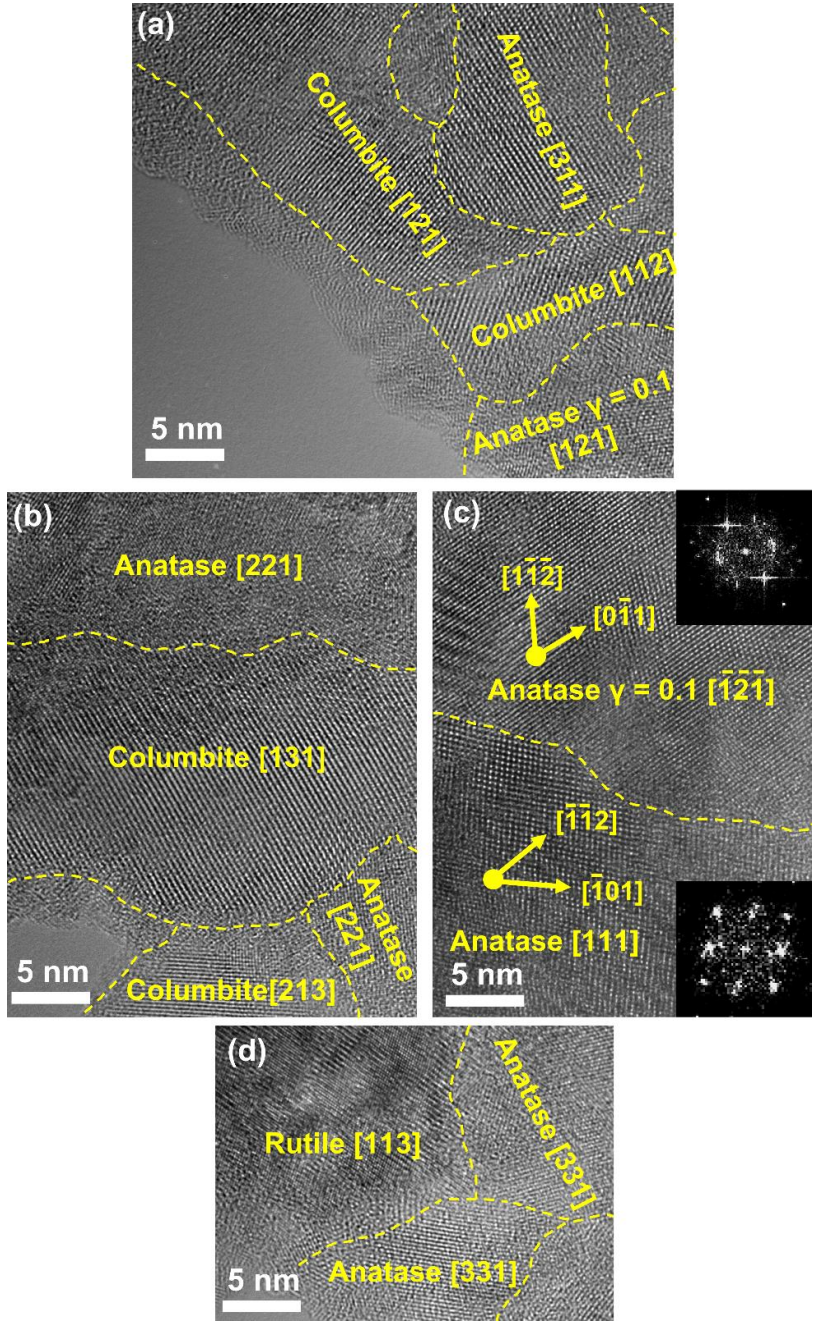

Fig. 7. High-resolution TEM images for $TiO_2$ sample processed by HPT under 6 GPa at room temperature for 15 turns.

## 4. Discussion

This study suggests a possible mechanism for the effect of shear strain on pressure-induced phase transformations through the formation of intermediate sheared phases. Such a mechanism can explain the acceleration of phase transformations by shear strain reported during HPT processing of various materials including metals such as cobalt, titanium, zirconium and their alloys [38-42], composites [44, 45], semiconductors [46, 47], ceramics [14-18], and high-entropy alloys and ceramics [48,49]. It was shown earlier that HPT processing not only refines the microstructure but also accelerates various allotropic phase transformations [11-13]. Specifically in ceramics, HPT was reported to significantly accelerate phase transformations by the shear strain effect [2,6]. In addition to accelerated phase transformations, the phases formed by HPT can usually remain metastable after processing for a long time [2,6]. For example, the high-pressure columbite phase was shown to remain stable in a wide range of temperatures (at least up to 773 K) and for a long period of time (at least three years) under ambient pressure after HPT processing [18]. Understanding the mechanisms of accelerated phase transformation and the stability of these phases can be fundamental for predicting the behavior of new materials under processing by HPT and other severe plastic deformation methods [1,2]. This study attempts to answer one main question: what is the possible mechanism for the effect of the shear strain on the anatase-to-columbite phase transformation during HPT?

To answer this question, one should consider that shear strain in HPT has four well-known contributions to pressure-induced phase transformations. First, the continuous production of defects (especially dislocations) during the process provides a pressure concentrator that reduces the threshold external pressure for phase transformations [18,50]. This behavior has been deeply investigated for BN ceramics by rotational diamond anvil cells [51, 52]. Second, the generation of defects can accelerate the atomic diffusion for phase transformation [11]. Such an effect was particularly reported at extremely large strains (i.e. after ultra-severe plastic deformation) [53]. Third, shear strain reduces the crystal size and influences the phase transformations by changing the phase stability through the size effect [18]. For example, it was reported that columbite remains stable for grain sizes smaller than 15 nm due to the competition between the bulk and interphase energies [18]. Fourth, in nanograined materials, the effect of external shear strain is enhanced at grain boundaries due to an additional strain effect in boundaries through atomic rearrangements [51]. This dynamic strain/boundary interaction contributes to overcoming the energy barrier for phase transformations [51]. In addition to these three contributions, the external shear strain may lead to strain in unit cells and produce intermediate sheared phases that can be formed with less energy barrier compared to the final high-pressure phases. This mechanism, which was examined in this study, can provide another justification for the acceleration of pressure-induced phase transformations by shearing because it is well-known that sheared phases such as martensite can be formed more easily in the presence of shear strain [54,55].

The proposed mechanism in this study suggests that phase transformations proceed through intermediate sheared phases that are a result of plastic flow within the material during HPT processing. These intermediate phases are less thermodynamically stable than the final high-pressure phases according to the quantum mechanics calculations, but the energy barrier for their transformation can be easily overcome by the external shear strain effect. Therefore, the mechanism suggests that the anatase-to-columbite phase transformation can follow a pathway such as pristine anatase → sheared anatase → sheared columbite → columbite, although sheared columbite could not be detected by TEM. This pathway, which was suggested based on first-principles energy and phonon calculations, was partly supported by the experiments because sheared anatase could be detected by carefully analyzing the XRD patterns and high-resolution micrographs. The unsuccessful attempt to find the dynamically stable sheared columbite phase suggests that either the fraction of this sheared phase is so small (because of its quick conversion to columbite) or it does not form as an intermediate phase during the transformation. Regardless of the exact transition path, which needs to be examined by further theoretical and experimental studies, the suggested mechanism based on intermediate phases can provide another insight into the significance of shear strain on pressure-induced phase transformations reported in a wide range of materials [11-13]. Moreover, such intermediate phases can be of interest from the property and application points of view, if they can be stabilized in a large fraction [30,31].

## 5. Conclusion

The combination of quantum mechanics calculations and experimentations using the high-pressure torsion method led to a new comprehension of the impact of shear strain in the anatase-to-columbite phase transformation. The experimental findings validate a mechanism of phase transformation that involves at least a sheared anatase phase as an intermediate phase. The suggested mechanism, which was examined for TiO$_2$, may contribute to other phase transformations reported during severe plastic deformation.

**Acknowledgments**

The author JHJ acknowledges a scholarship from the Q-Energy Innovator Fellowship of Kyushu University. This study is supported partly by Mitsui Chemicals, Inc., Japan, partly through Grants-in-Aid from the Japan Society for the Promotion of Science (JP19H05176, JP21H00150 & JP22K18737), and partly by Japan Science and Technology Agency (JST), the Establishment of University Fellowships Towards the Creation of Science Technology Innovation (JPMJFS2132).

**Author Contributions**

All authors contributed to conceptualization, methodology, validation, investigation and writing original draft.

**Conflict of Interest**

The authors declare no conflicts of interest or no competing interests:

**Data and Code Availability**

The data can be made available on request.

**Supplementary Information**

Not applicable

**Ethical Approval**

This study did not use human or animal tissues.